\documentclass[12pt]{iopart}
\usepackage{graphicx}
\newcommand{\beq}{\begin{eqnarray}}
\newcommand{\eeq}{\end{eqnarray}}

\newcommand{\bq}{\vec{q}}

\newcommand{\bk}{\vec{k}}
\newcommand{\brr}{\vec{r}}
\newcommand{\bx}{\vec{x}}
\newcommand{\by}{\vec{y}}

\newcommand{\kslash}{k\kern-1ex /}
\newcommand{\qslash}{q\kern-1ex /}
\newcommand{\lslash}{l\kern-1ex /}
\newcommand{\sslash}{s\kern-1ex /}
\newcommand{\paslash}{p_a\kern-2ex /}
\newcommand{\pbslash}{p_b\kern-2ex /}
\newcommand{\Dslash}{{\cal D}\kern-1.5ex /}
\newcommand{\dslash}{\partial\kern-1.2ex /}

\begin{document}

\begin{flushright}
UTHEP-560, UTCCS-P-42\\
TKYNT-08-06
\end{flushright}

\title[Nuclear Force from Lattice QCD]{Nuclear Force  from Monte Carlo Simulations of  \\
Lattice Quantum Chromodynamics}

\author{S. Aoki}
\address{Graduate School of Pure and Applied Sciences,
	 University of Tsukuba, \\
	 1-1-1 Tennodai, Tsukuba 305-8571, JAPAN \\
	 RIKEN BNL Research Center, Brookhaven National Laboratory, \\
	 Upton, New York 11973, USA}
\ead{saoki@het.ph.tsukuba.ac.jp}

\author{T. Hatsuda}
\address{Department of Physics, The University of Tokyo, \\
	7-3-1 Hongo, Bunkyo-ku, Tokyo 113-0033, JAPAN}
\ead{hatsuda@phys.s.u-tokyo.ac.jp}

\author{N. Ishii}
\address{Center for Computational Sciences,
	University of Tsukuba, \\
	1-1-1 Tennodai, Tsukuba 305-8571, JAPAN}
\ead{ishii@rarfaxp.riken.jp}

\begin{abstract}

 The nuclear force acting between protons and neutrons
  is studied in the Monte Carlo simulations of the fundamental theory
  of the strong interaction, the quantum chromodynamics defined 
  on the hypercubic space-time lattice.
   After a brief summary of the empirical nucleon-nucleon
  (NN) potentials which can fit the NN scattering experiments in high precision,
 we outline the basic formulation to 
  derive the potential between the extended objects such as 
  the nucleons composed of quarks. 
  The equal-time Bethe-Salpeter amplitude
  is a key ingredient for defining the NN potential on the lattice.
  We show the results of the numerical simulations
  on a $32^4$ lattice with the lattice spacing $a \simeq 0.137 $fm 
  (lattice volume (4.4 fm)$^4$)
   in the quenched approximation.
 The calculation was carried out  using the massively parallel computer
  Blue Gene/L at KEK.
  We found that
  the  calculated NN potential at low energy has basic features 
  expected from  the empirical NN potentials; attraction
   at  long and medium distances and the repulsive core at short
  distance.  Various future directions along this line of research
  are also summarized. 

\end{abstract}


\maketitle
\section{Introduction}
  
 One of the long standing problems
 in particle and nuclear physics is the
  origin of the strong nuclear force which holds the nucleons (protons and neutrons)
  inside atomic nuclei.   For the past half century, 
   phenomenological fits of the  proton-proton (pp) and 
  neutron-proton (np) scattering data assuming 
  empirical  nucleon-nucleon (NN)
  potentials have been attempted  \cite{NN-review,NN-data}:
 The potentials, which can fit
 more than 2000 data points of the NN phase shift with $\chi^2/{\rm dof} \simeq 1$  
 for $T_{\rm lab} < 300 $MeV, include the CD-Bonn potential \cite{CD-Bonn},
 Argonne $v_{18}$ potential \cite{Wiringa:1994wb}
 and Nijmegen potentials \cite{Stoks:1994wp}.
 Alternative approach on the basis of the 
  chiral perturbation theory has been also developed \cite{ChPT}.
 
\begin{figure}[t]
\begin{center}
\includegraphics[width=8cm]{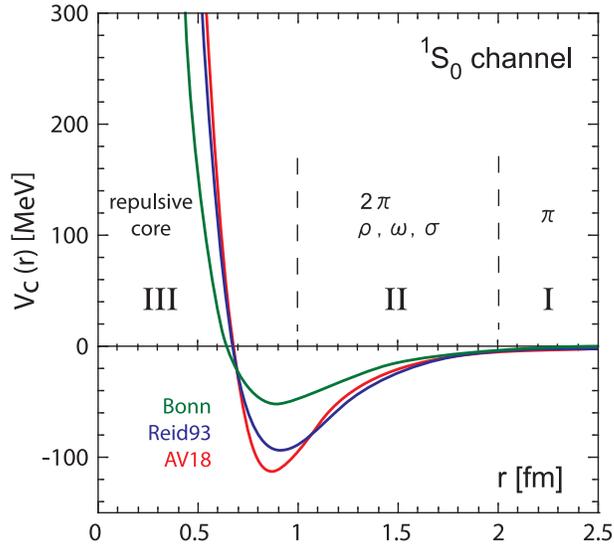}
\end{center}
\caption{Three examples of the high-precision 
NN potentials in the $^1S_0$ 
channel. AV18 stands for the Argonne $v_{18}$ potential \cite{Wiringa:1994wb}, 
 Reid93 stands for one of the Nijmegen potentials \cite{Stoks:1994wp}
 and Bonn stands for the Bonn potential \cite{Bonn}.
I, II and III correspond to the long range part, medium range part and 
the short range part, respectively, as discussed in the text.  
 
 }
\label{phen-pot}
\end{figure}

  Shown in  Fig.\ref{phen-pot} are three examples of
  the empirical NN potentials in the $^1S_0$ channel.\footnote{A system of two nucleons with
   total spin $s = (0, 1) $, orbital angular momentum $L =(S, P, D, \cdots)$ and total angular momentum 
   $J = (0, 1, 2, \cdots)$ is   denoted as  $^{2s+1} L_J$.} 
    From this figure, some  characteristic features of the nuclear force can be seen: 
 \begin{enumerate}
  \item[I.] The long range part of the nuclear force  ($r >  2$  fm) 
  is dominated by the one pion exchange originally introduced by 
  Yukawa \cite{yukawa}.   
  \item[II.] The medium range part ($1\ {\rm fm} <  r < 2$ fm) receives 
  significant contributions from the exchange of
  multi-pions and heavy mesons ($\rho$, $\omega$, and $\sigma$).
  In particular, the spin-isospin independent attraction
  of about $-50 \sim -100$ MeV in this region plays an essential role
  for binding  the atomic nuclei.
 \item[III.] The short  range part ($r < 1$ fm) behaves as 
  a repulsive core originally introduced by Jastrow \cite{jastrow}
  to explain the pp and np scattering phase shifts simultaneously. 
  Such a short range repulsion is relevant to  
   the stability of atomic nuclei, 
  to  the maximum mass of neutron stars, and to 
 the ignition of the Type II supernova explosions \cite{VJ}.
\end{enumerate}

 It is now well established that the nucleons  are made of
  quarks and gluons which obey the law of quantum
   chromodynamcs (QCD) \cite{wilczek05}. Therefore, it is tempting to derive the 
   strong nuclear force from the quark-gluon degrees of freedom.
     Two nucleons have a sizable overlap
  at  distance $r < 1$ fm since the proton's  charge radius is
  about 0.86 fm. Therefore, the nuclear force at short
   distances must  be described by taking into account
   the  direct exchange of quarks  and gluons between the nucleons.
  So far,  there have been numerous theoretical attempts to understand the
  nuclear force from the quark structure
   of the nucleon \cite{QQ_review}. However, conclusive
  results have not been obtained because of the highly 
  non-perturbative  nature of QCD.
 
 Recently,  the present authors have developed a new approach to this
 long standing problem of nuclear force 
 on the basis of QCD defined 
 on a space-time lattice (LQCD) \cite{wilson,lat07}.
  In LQCD, physical quantities are expressed
  by  highly multi-dimensional integrals  which can be carried out 
  by  importance sampling method.
 Our idea is  to define the NN potential in the coordinate space from the 
  equal-time  Bethe-Salpeter amplitude calculated on the lattice:
   The first results   on the central NN potential in the 
  $^1S_0$ and $^3S_1$ channels have been reported   
  with the quenched lattice QCD simulations in \cite{Ishii_PRL,Ishii_lat07,aoki_lat07}.
 Also the first results of the hyperon-nucleon potential have
 been reported in \cite{Nemura_lat07,Nemura08}.

  In the following,
  we  will outline the field theoretical derivation of the NN potential
   from QCD \cite{AHI_full}
  in Sec.\ref{sec:BS-wave} and Sec.\ref{sec:NN-pot}.
 Then, we show how to define  the potential in 
  LQCD formalism in 
  Sec.\ref{sec:effective-C}.  The basic setup and the method of our numerical
  simulations are shown in Sec.\ref{sec:setup}.  Some  
  numerical results of the low energy NN potential taken
   from \cite{Ishii_PRL,Ishii_lat07} are shown in Sec.\ref{sec:numerical}.
   The last section is devoted to summary and concluding remarks with 
   a discussion on the future directions.

\section{Bethe-Salpeter wave function and the NN potential}
\label{sec:BS-wave}

Let us start with the definition of the Bethe-Salpeter (BS) amplitude
 for the proton-neutron system,
\begin{eqnarray}
\Psi_{\alpha \beta}(x,y)  &=& \langle 0 \vert 
{\rm T}[ \hat{n}_{\beta}(y) \hat{p}_{\alpha}(x) ] \vert 
{\rm p}(\bq, s) {\rm n}(\bq\ ', s') ;{\rm in} \rangle  , 
  \label{BS-amp} \\
   \hat{n}_\beta(y) &=&
  \varepsilon_{abc} \left(
		     \hat{u}_a(y) C \gamma_5 \hat{d}_b(y)
		    \right)
  \hat{d}_{c\beta}(y),
  \label{neutron} \\
    \hat{p}_\alpha(x) &=&
  \varepsilon_{abc} \left(
		     \hat{u}_a(x) C \gamma_5 \hat{d}_b(x)
		    \right)
  \hat{u}_{c\alpha}(x),
  \label{proton}
\label{eq:BS-def}
\end{eqnarray}   
where $(\bq, s)$ and $(\bq\ ', s')$ denote the 
 spatial momentum and the spin-state of the incoming proton and those of 
 the neutron, respectively. The local composite operator for the 
  neutron (proton) are denoted by 
  $\hat{n}_{\beta}(y)$ ($\hat{p}_{\alpha}(x)$) with the operators for the up-quark
   $\hat{u}(x)$ and the down-quark $\hat{d}(x)$. Also, 
  $\alpha$ and $\beta$ denote the Dirac indices, 
$a$, $b$ and $c$ the color indices, and 
$C$ the charge conjugation matrix in the spinor space.
  
   One of the advantages to use 
   local operator for the nucleon 
   is that the  Nishijima and Zimmermann (NZ) reduction formula \cite{NZ} 
   for local composite fields can be utilized.  
   In particular, in and out nucleon fields are
   defined through the Yang-Feldman equation,
 \beq
 \sqrt{Z} \hat{N}_{\rm in/out}(x) 
 = \hat{N}(x) - \int S_{\rm adv/ret}(x-x';m_{\rm N}) \hat{J}(x') dx'  ,  
\label{YF-eq}
 \eeq
 where $\hat{N}(x)$ is the nucleon composite operator  
 ($\hat{n}(x)$ or $\hat{p}(x)$ in Eqs.(\ref{neutron},\ref{proton})),  
  $\hat{N}_{\rm in/out}(x)$ is the associated in/out field,
   $m_{\rm N}$ is the physical nucleon mass, and  $\hat{J}(x)
  \equiv (i \dslash_x - m_{\rm N}) \hat{N}(x)$.
  The advanced/retarded propagator for the free nucleon field with the 
   mass $m_{\rm N}$ is denoted by $S_{\rm adv/ret}(x-x';m_{\rm N})$.
   The  normalization factor, $\sqrt{Z}$,
  is the coupling strength of the composite operator $\hat{N}(x)$
  to the physical nucleon state. 
 
   Through the NZ   reduction formula, 
   the BS amplitude in Eq.(\ref{BS-amp})  is related to the 
   four-point  Green's function $G_4$ of the composite nucleons which is
    decomposed into the free part and the scattering part, 
    $G_4= Z^2 ( G_4^{(0)} + G _4^{\rm (sc)})$, where $G_4^{(0)}$ is
    proportional to a product of the free nucleon propagators.
   After taking  
   the equal time limit, $x^0=y^0=t$, with $\brr \equiv \bx -\by $, 
   it is straight forward to 
   rewrite Eq.(\ref{BS-amp}) to the  the following
    integral equation in the c.m. frame ($\bq\ '=-\bq$) 
  \cite{AHI_full},
  \beq
\label{eq:LS-eq1}
  \Psi_{\alpha \beta}(\brr, t)&=&
   \psi_{\alpha \beta}(\brr; \bq,s,s')  e^{-2i \sqrt{\bq^2 + m_{\rm N}^2} t},\\
\! \! \! \! \! \!  \! \! \!  \!  \! \! \!  \psi_{\alpha \beta}(\brr; \bq,s,s') &=&
   Z u_{\alpha}(\bq,s) u_{\beta} (- \bq, s') e^{i \bq \cdot \brr} \nonumber \\  
 & &  \ \ \ \ \ \  +  Z \sum_{\gamma \delta}
 \int \frac{d^3k}{(2\pi)^3} \ e^{i \bk \cdot \brr}
  {\cal F}_{\alpha \beta; \gamma \delta}(\bk ; \bq) 
  u_{\gamma}(\bq,s) u_{\delta} (- \bq, s').
\label{eq:LS-eq2}
   \eeq 
 Here   $u_{\alpha}(\bq;s)$ is the positive-energy plain-wave solution of the   
 Dirac equation and  $  {\cal F}_{\alpha \beta; \gamma \delta}(\bk ; \bq)$
 is an integral kernel obtained from $G_4^{\rm (sc)}$ after
   carrying out the $k_0$-integration.  
 Hereafter, we call $\psi_{\alpha \beta}(\brr; \bq,s,s')$ as the Bethe-Salpeter
  wave function. 
   The differential form of the above equation 
 is obtained by multiplying $(\bq\ ^2 + \nabla^2)/m_{\rm N}$ 
 to Eq.(\ref{eq:LS-eq2});
 \beq
 \label{eq:KG-eq}
\frac{1}{m_{\rm N}}
(\bq^{\ 2} + \nabla^2) \psi_{\alpha \beta}(\brr; \bq,s,s') 
= K_{\alpha \beta}(\brr; \bq,s,s').
 \eeq  
  Note that we have not made any non-relativistic approximation
  to derive  Eq.(\ref{eq:KG-eq}).
 An important observation  is that
 the plain wave component of $ \psi_{\alpha \beta}(\brr; \bq,s,s') $
 is projected out by the operator $(\bq^{\ 2} + \nabla^2)$ so that 
 the function $K_{\alpha \beta}(\brr; \bq,s,s')$ is localized in 
 coordinate space as long as $|\bq |$ stays below the inelastic
  threshold as noted for pion-pion scattering in \cite{Lin:2001ek}.
 This is equivalently said that the Fourier transform of 
 $K$ with respect to $\brr$, which is proportional to the 
  half off-shell $T$-matrix relating the 
  the on-shell state with momentum $\bq$ and the 
   off-shell state with  momentum $\bk$, does not develop a real pole as a 
   function of $|\bk|$, if $|\bq|$ is below
   the inelastic threshold.

\section{The NN potential}
\label{sec:NN-pot}
 
 In an abbreviated notation, $(\brr, \alpha, \beta) \rightarrow x$,
 and $(\bq, s, s') \rightarrow q$, Eq.(\ref{eq:KG-eq}) is written as
 \beq
 (E_q-H_0) \psi(x;q) =  K (x;q) ,
\label{eq:KGK-eq}
 \eeq
 where $E_q=\bq^2/m_{\rm N}$ and $H_0=-\nabla^2/m_{\rm N}$.
  This equation defined in a finite box  can be used
   to extract various information on the NN scattering from the 
    lattice QCD simulations:
 \begin{enumerate}
  \item[(i)]  Consider $K(x;q)$ as a measure to identify the 
   length $R$ beyond which the two nucleons do not 
    interact.  If we stay in such a region 
    where $K(x>R;q) \simeq 0$,  the wave function 
   $\psi(x>R;q)$ can be expanded by the solution of the Helmholtz
    equation inside a finite box. Then one can extract the phase shift given
   the incoming energy $E_q$.   This is the approach originally
    proposed in \cite{luscher} and is later
     elaborated to study hadron-hadron scatterings
     on the lattice \cite{kuramashi95,Lin:2001ek,ishizuka05,savage06}.  
 \item[(ii)] 
  Alternatively,  one may extract the half off-shell $T$-matrix in momentum space
   by  calculating the left-hand-side of   Eq.(\ref{eq:KGK-eq}) in the
    coordinate space and making Fourier transform with respect to $x$.
\item[(iii)] One can go one-step further   
 and define the non-local NN potential $U(x,x')$ 
 from  $K(x;q)$, so that 
 Eq.(\ref{eq:KGK-eq}) becomes the Schr\"{o}dinger type equation. 
 \end{enumerate}

 If we are interested only in the NN scattering phase shift in the free space,
 the procedure (i) is certainly enough.
 On the other hand, if we are interested in applying 
  Eq.(\ref{eq:KGK-eq}) to the problems of 
  bound states and the nuclear many-body system, 
 (ii) and (iii) are useful since they give us the off-shell information
 in a well-defined manner in QCD.  To see this explicitly for the case
  (iii),  we introduce  a set of functions labeled by $q$, 
  $\{ \tilde{\psi}(x;q) \}$, which is dual to the set $\{ {\psi}(x;q) \}$ in the 
  following sense:
 \beq
 \int dx \   \tilde{\psi}(x;q) \psi(x;p) = \delta_{q,p}.
\label{eq:dual-psi}
 \eeq
As long as the dimensions of the $x$-space and $p$-space
 are the same and  the elements in $\{ \psi(x;p) \}$ are   
 linearly independent, such a dual basis exists and is 
 unique.
 If the dimension of $p$-space is less than that of $x$-space,
 the dual basis exists but is 
 not unique.\footnote{
 This is easily seen as follows. Let
 us introduce a basis $\{ {\bf e}_{(1)}, {\bf e}_{(2)}, \cdots, {\bf e}_{(N)} \}$
  in the $N$-dimensional vector space.
 The BS wave function in discretized coordinates $ \psi(x;p) \equiv \psi(i;\alpha)$
 corresponds to  ${\rm e}^i_{(\alpha)}$ with $1 \le i \le N$ and $1 \le \alpha \le M
  \le N$.  If $M=N$, there exits a unique dual basis, 
 $\{ \tilde{\bf e}_{(1)}, \tilde{\bf e}_{(2)}, \cdots, \tilde{\bf e}_{(N)} \}$
 satisfying 
  $\tilde{\bf e}_{(\alpha)} \cdot {\bf e}_{(\beta)} = \delta_{\alpha \beta}$
 (see any textbook of linear algebra).  If $M < N$,   
  there is still a dual basis
    $\{ \tilde{\bf e}_{(1)}, \tilde{\bf e}_{(2)}, \cdots, \tilde{\bf e}_{(M)}\} $
    satisfying the above condition for
   $\alpha \le M$ and $ \beta \le M$.
    However, it is not unique because one always has a freedom
  to add linear combinations of $\tilde{\bf e}_{\gamma}$ ($M+1 \le \gamma \le N$)
    to the  above dual basis. }
   Assuming the existence (but not necessarily the uniqueness)
  of the dual basis, the non-local potential can be defined
  as
 \beq 
 \label{eq:U-def}
 U(x,x') = \int dp \ K(x,p) \tilde{\psi}(x';p).
 \eeq
 The Eqs.(\ref{eq:dual-psi},\ref{eq:U-def})
  lead to the formula, $K(x;q) = \int dx' \  U(x,x') \psi(x';q)$,
 so that  Eq.(\ref{eq:KGK-eq}) becomes
 \beq
 \! \! \! \!  \! \! \! \!  \! \! \! \!  \! \! \! \!  \! \! \! \!  \! \! \! \! 
 \frac{-\nabla^2}{m_{\rm N}}  \psi_{\alpha \beta}(\brr;\bq,s,s')  
 + \int d^3r'  \ U_{\alpha \beta; \gamma \delta} (\brr,\brr\ ') 
 \psi_{\gamma \delta}(\brr\ ';\bq, s, s' ) 
  = E_q \psi_{\alpha \beta}(\brr: \bq, s, s').
 \eeq

Note that the non-local potential $U$ can be  
rewritten in the form,
\begin{equation}
U(\brr,\brr\ ')=
 V(\brr,\nabla)\delta(\brr-\brr\ ').
\end{equation}
The general structure of $V(\brr,\nabla)$ under various
 symmetry constraints in the non-relativistic
 kinematics  has been worked out by Okubo and Marshak \cite{okubo}.
 If we further make the derivative expansion at low 
 energies \cite{TW67},  we obtain  the expression familiar
 in the phenomenological potentials acting on the upper components of 
 the wave function;
\begin{eqnarray}
 \! \! \! \!  \! \! \! \!  \! \! \! \!  \! \! \! \!  \! \! \! \!  \! \! \! \! 
 \!  \! \! \! \! 
 V(\brr,\nabla)   & =&
  V_0(r)
  +V_\sigma(r)(\vec{\sigma}_1\cdot\vec{\sigma}_2)
  +V_\tau(r)(\vec{\tau}_1\cdot\vec{\tau}_2)
  +V_{\sigma\tau}(r)
   (\vec{\sigma}_1\cdot\vec{\sigma}_2)
   (\vec{\tau}_1\cdot\vec{\tau}_2)
   \nonumber \\
 &&
  +V_{\rm T}(r)S_{12}
  +V_{{\rm T}\tau}(r)S_{12}(\vec{\tau}_1\cdot\vec{\tau}_2)
  +V_{\rm LS}(r)(\vec{L}\cdot\vec{S})
  +V_{{\rm LS}\tau}(r)(\vec{L}\cdot\vec{S})(\vec{\tau}_1\cdot\vec{\tau}_2)
  \nonumber \\
 &&
  +{O}(\nabla^2). 
\label{eq:OM-decom}
\end{eqnarray}
Here
$S_{12}=3(\vec{\sigma}_1\cdot\vec{n})(\vec{\sigma}_2\cdot\vec{n})-\vec{\sigma}_1\cdot\vec{\sigma}_2$
is the tensor operator with $\vec{n}=\brr/|\brr|$,
$\vec{S}=(\vec{\sigma}_1 + \vec{\sigma}_2)/2$ 
the total spin operators,
$\vec{L}=-i\brr\times {\nabla}$ the orbital 
angular momentum operator, 
and $\vec{\tau}_{1,2}$ are the isospin operators 
for the nucleons.  
 Each component of the potential in Eq.(\ref{eq:OM-decom})
 can be obtained by appropriate spin, isospin and angular momentum
  projection of the BS wave function. Also, the higher
   derivative terms of the potential in Eq.(\ref{eq:OM-decom})
    can be deduced  by combining the BS wave functions
  for different incident energies.

 It is in order here to remark that the structure of the non-local potential $U(x,x')$
 is directly related to the nucleon interpolating operator 
 adopted in defining the Bethe-Salpeter wave function.
 Different choices of the interpolating operator would
 give  different forms of the NN potential 
 at short distance, although they give the same 
 phase shift at asymptotic large distance.
 The advantage of working in QCD is that
  we can unambiguously trace the connection between the NN potential
   and the interpolating    operator.

\section{Effective central potential on the lattice}
\label{sec:effective-C}

 The BS wave function in the $S$-wave on the lattice with the 
 lattice spacing $a$ and the spatial lattice volume $L^3$ is obtained  by
\begin{eqnarray}
 \psi({r}) &=&
  {1\over 24} \sum_{{ R}\in O} {1\over L^3} \sum_{\vec{x}}
  P^\sigma_{\alpha\beta} 
  \left\langle 0
   \left|
    \hat{n}_\beta({ R}[\vec{r}]+\vec{x})
    \hat{p}_\alpha(\vec{x})
   \right| {\rm pn}  ; q
  \right\rangle .
 \end{eqnarray}
The summation over ${ R}\in O$ is taken for the cubic transformation 
group to project out the $S$-wave.\footnote{
 In principle, this projection cannot remove possible 
 contamination from the higher orbital waves with $L \ge 4$, although
  these contributions are expected to be negligible.}
The summation over $\vec{x}$ is to select the state with  zero total momentum. 
We take the upper components of the Dirac indices 
to construct the spin singlet (triplet) channel by 
$P^{\sigma={\rm singlet}}_{\alpha\beta}=(\sigma_2)_{\alpha\beta}$
($P^{\sigma={\rm triplet}}_{\alpha\beta}=(\sigma_1)_{\alpha\beta}$). 
The BS wave function $\psi(\vec{r})$ 
is understood as the probability amplitude to find 
``neutron-like'' three-quarks located at point $\vec{x}+\vec{r}$ and
``proton-like'' three-quarks located at point $\vec{x}$.
  
In the actual simulations, the BS wave function is obtained from the 
four-point correlator,
\begin{eqnarray}
\label{eq:4-point}
\! \! \! \! 
\! \! \! \! \! \! \! \! \! \! \! \! \! \! \! \! \! \! \! \! \! \! \! \! \! \! \! \! \! \! \! \! 
 G_4(\vec{x}, \vec{y}, t; t_0) &=&
  \left\langle 0
   \left|
    \hat{n}_\beta(\vec{y},t)
    \hat{p}_\alpha(\vec{x},t)
    \overline{\cal J}_{pn}(t_0)
   \right| 0 
  \right\rangle
  = 
  \sum_n A_n
  \left\langle 0
   \left|
    \hat{n}_\beta(\vec{y})
    \hat{p}_\alpha(\vec{x})
    \right| n
  \right\rangle
  {\rm e}^{-E_n(t-t_0)}.
\label{eq:BSamp}
\end{eqnarray}
Here $\overline{\cal J}_{pn}(t_0)$ is a wall source located at
$t=t_0$,
which is defined by 
${\cal J}_{pn}(t_0)=
P_{\alpha\beta}^{\sigma} \sum_{\vec{x},\vec{y}}
\hat{p}_{\alpha}(\vec{x},t_0)
\hat{n}_{\beta}(\vec{y},t_0)$.
 The eigen-energy and the eigen-state of the
  six quark system are denoted by $E_n$ and $|n\rangle$, respectively,
  with the matrix element $A_n(t_0)=\langle n|\overline{\cal J}_{pn}(t_0)|0\rangle$. 
For $(t-t_0)/a \gg 1$,
  the $G_4$ and hence the wave function $\psi$ are dominated 
by the lowest energy state. 

The lowest energy state 
 created by the wall source $\overline{\cal J}_{pn}(t_0)$
 contains not only the $S$-wave component but also 
 the $D$-wave component induced by  the tensor force.
 In principle, they can be disentangled
 by preparing appropriate operator sets for the sink.
 Study along this line to extract the mixing between the
  $S$-wave and the $D$-wave at low energies
  has been put forward recently in \cite{Ishii_chiral07}. 
   In the present paper, instead of making such decomposition, we 
 define  an ``effective" central potential $V_{\rm C}(r)$
  according to Refs.\cite{Ishii_PRL,Ishii_lat07}: 
  \beq
  V_{\rm C}(r) = 
  E_q + {1\over m_{\rm N}}{{\nabla}^2\psi(r)\over \psi(r)} .
\label{eq:naive_pot}
\eeq
 Note that one can 
 test the non-locality of the potential $U(x,x')$ by evaluating 
 the effective central potential for different energies.
  If there arises appreciable energy dependence 
  in  $V_{\rm C}(r)$, it is a signature of the 
  necessity of high derivative terms in 
  Eq.(\ref{eq:OM-decom}).

\section{Setup of the lattice simulations}
\label{sec:setup}

In lattice QCD simulations, the vacuum expectation value of 
an operator ${\cal O}(q,\bar{q},U)$ is defined as
\beq
\langle {\cal O} \rangle 
&=& {\cal Z}^{-1} \int \prod_\ell dU(\ell) 
\prod_x dq(x) d\bar{q}(x) \ {\cal O}(q,\bar{q},U) \ e^{-S_{\rm f}(q,\bar{q},U) - S_{\rm g}(U)}  \label{eq:full_QCD}\\
&=& {\cal Z}^{-1} \int \prod_\ell dU(\ell)  \ Q(U) \det M(U)\  e^{-S_{\rm g}(U)}. \label{eq:quenched_QCD}
\eeq
where ${\cal Z}=\int \prod_\ell dU(\ell) \prod_x dq(x) d\bar{q}(x) \  e^{-S_{\rm f}(q,\bar{q},U) - S_{\rm g}(U)} $
 is the QCD partition function, 
$S_{\rm f} = \sum_{x,x'} \bar{q}(x) M_{xx'}(U) q(x')$ is the quark part of the action, and
 $S_{\rm g}$ is the gluon part of the action.
 The quark field $q(x)$ is defined on each site $x$ of the hypercubic space-time lattice, while
the gluon field $U(\ell)$ denoted by $3 \times 3$ special unitary matrix  is defined on each link
 $\ell$. In Eq.(\ref{eq:quenched_QCD}), the integration of the quark fields is 
 carried out analytically.  
   In the quenched approximation adopted in our simulation, the virtual fermion loop
 denoted by $\det M(U)$ is set to be 1 and the integration over
  link variables $U$ is performed using the importance sampling method \cite{LQCD-text}.
  
    We  employ  the  standard  plaquette  gauge action on  a $32^4$  lattice  with
 the bare QCD coupling constant $\beta = 6/g^2 =  5.7$.
 The corresponding lattice spacing  is determined to be 
  $1/a=1.44(2)$ GeV  ($a\simeq 0.137$ fm)  
  from the $\rho$ meson mass in the chiral limit \cite{kuramashi96}.
 Then, the physical size of our lattice becomes $L\simeq 4.4$ fm.
   For the fermion action, we adopt the standard Wilson quark action  with the hopping parameter
 ($\kappa=0.1640, 0.1665$ and $0.1678$) which controls the quark masses.
 The  periodic (Dirichlet) boundary  condition is  imposed on  the quark
fields along the spatial (temporal) direction.
  To generate the quenched gauge configurations,  
  we adopt the heatbath algorithm with overrelaxation and
 sample configurations are taken in every 200 sweeps after skipping
 3000 sweeps for thermalization.

For our numerical simulations,  we use IBM  Blue Gene/L at
KEK, which consists of 10,240 computation nodes with total theoretical
performance of 57.3 TFlops.
A modified version of the CPS++ (the Columbia Physics System) \cite{cps}
is used  to generate quenched gauge configurations  and propagators of
quarks.
Most  of the  computational  time  is devoted  to  the calculation of
the four-point function of  nucleons, for which our code  achieves 34--48 \%
of peak performance.
Totally  about 4000  hours are  used by  queues with  512 nodes  for the
calculations of the effective  central potentials corresponding to the
three values of hopping parameters.
The  number of  sampled gauge configurations  $N_{\rm conf}$,  the  pion mass
$m_{\pi}$,   and the  nucleon   mass   $m_{\rm  N}$   are  summarized   in
Table~\ref{table}.  (For $\kappa=0.1678$,  we have removed 
24 exceptional gauge configurations from the sample.)
\begin{table}
\begin{center}
\begin{tabular}{cccccll}
\hline
$\kappa$ & $N_{\rm conf}$ & $m_{\pi}$ [MeV] & $m_{\rm N}$ [MeV] & $(t-t_0)/a$
& $E_q(^1S_0)$ [MeV] & $E_q(^3S_1)$ [MeV] \\
\hline
0.1640   & 1000           & 732.1(4) & 1558.4(63) & 7 & $-0.400(83)$ & $-0.480(97)$\\
0.1665   & 2000           & 529.0(4) & 1333.8(82) & 6 & $-0.509(94)$ & $-0.560(114)$\\
0.1678   & 2021           & 379.7(9) & 1196.6(32)& 5  & $-0.675(264)$ & $-0.968(374)$\\
\hline
\end{tabular}
\end{center}
\caption{The number  of gauge configurations $N_{\rm  conf}$, the pion
mass $m_{\pi}$,  the nucleon mass  $m_{\rm N}$, time-slice  $t-t_0$ on
which  BS  wave  functions  are  measured,  and  the  non-relativistic
energies $E_q \equiv q^2/m_{\rm N}$ for $^1S_0$ and $^3S_1$ channels.}
\label{table}
\end{table}

 We adopt the wall source
on the time-slice $t/a=t_0/a = 5$.  The BS wave functions are measured
on  the time-slice  $(t-t_0)/a  =  7, 6,  5$  for $\kappa=0.1640,  0.1646,
0.1678$,  respectively.  The  ground state  saturation is  examined by
the $t$-dependence  of  the NN  potential.   We  employ the  nearest
neighbor representation of  the discretized Laplacian as $\nabla^2
f(\vec x) \equiv \sum_{i=1}^{3}\left\{ f(\vec  x + a\vec n_i) + f(\vec
x -  a\vec n_i)\right\} - 6  f(\vec x)$, where $\vec  n_i$ denotes the
unit vector  along the $i$-th  coordinate axis. BS wave  functions are
fully measured for  $r <  0.7$ fm, where  rapid change of the
 NN potential is expected. 
Since  the  change  is  rather  modest for  $r >  0.7$
fm,   the
measurement of BS wave functions  is restricted on the coordinate axes
and their  nearest neighbors to  reduce the calculational  cost. 
The ``asymptotic momentum'' $q$ is  obtained 
 by fitting the BS wave function with the Green's function in a
finite and periodic box \cite{luscher}:
\begin{eqnarray}
\label{eq:GF}
 \! \! \!  \! \! \! G(\vec   r; q^2)
  & = &
  \frac1{L^3}
  \sum_{\vec n \in {\bf Z}^3}
    \frac{e^{ i (2\pi/L) \vec n \cdot\vec r }    }   { (2\pi/L)^2 {\vec n}^2 - q^2},  
\end{eqnarray}
which satisfies $(\nabla^2 + q^2) G(\vec   r; q^2) = - \delta_L(\vec r)$ with
 $\delta_L(\vec r)$ being the periodic delta-function.
 In the actual calculation,  Eq.(\ref{eq:GF}) is rewritten in terms of the 
 heat kernel ${\cal H}$ satisfying the heat equation,
  $\partial_t {\cal H}(t, \vec r) = \nabla^2 {\cal H}(t, \vec r)$ with 
   the initial condition, $ {\cal H} (t \rightarrow 0+, \vec r) = \delta_L (\vec r)$. 
  The fits are  performed outside the range  of NN interaction
   determined by  $\nabla^2\psi(\vec r)/\psi(\vec  r)$
\cite{ishizuka05}.

\section{Numerical results}
\label{sec:numerical} 
 
\begin{figure}[t]
\begin{center}
\includegraphics[width=7cm,angle=-90]{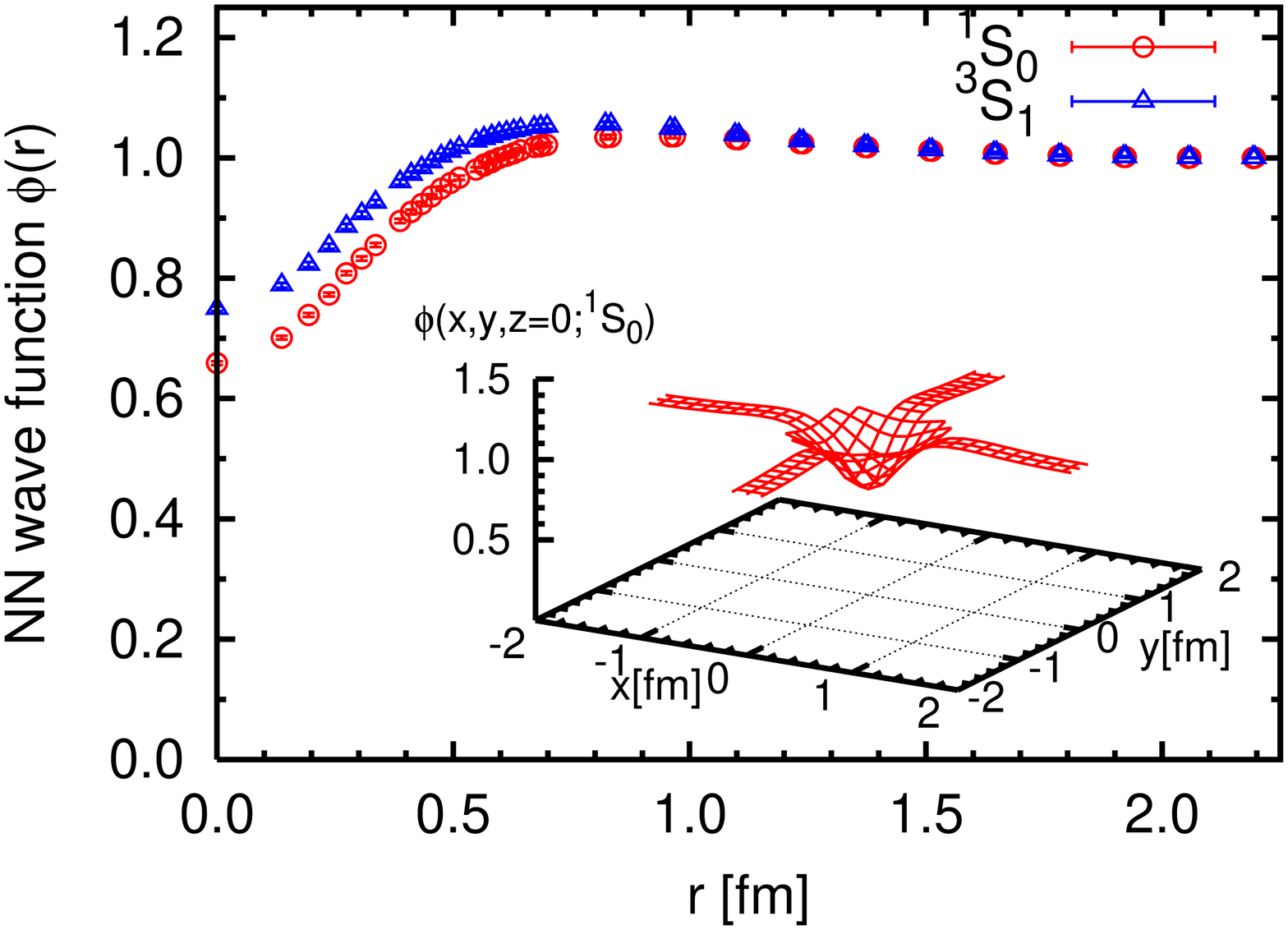}

\vspace{0.5cm}

\includegraphics[width=7cm,angle=-90]{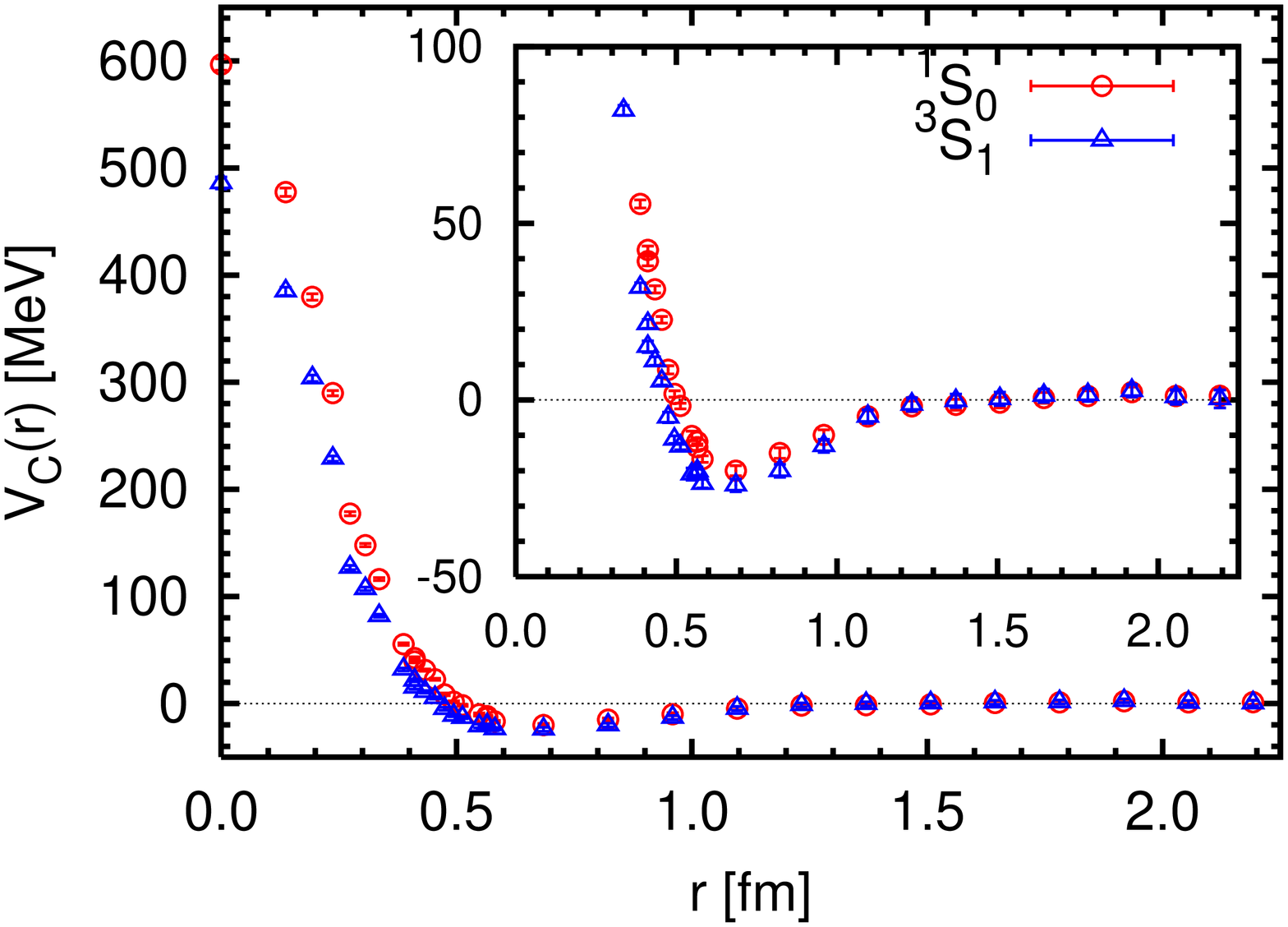}
\end{center}
\caption{(Upper panel) The NN wave functions in $^1S_0$ and $^3S_1$ channels.
The  inset  is a  3D  plot of  the wave  function $\phi(x,y,z=0;\  ^1S_0)$.
 (Lower panel) The NN  effective  central  potential  in the  $^1S_0$  and
  $^3S_1$ channels for $m_{\pi}=529$ MeV
 ($\kappa=0.1665$).
}
\label{fig1}
\end{figure}

\begin{figure}[t]
\begin{center}
\includegraphics[width=8.5cm,angle=-90]{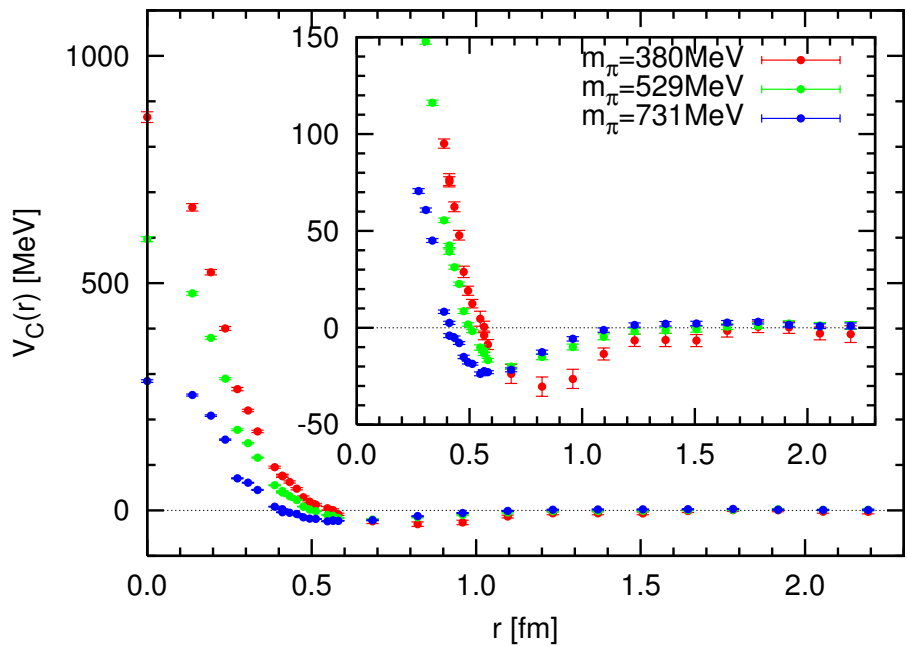}
\end{center}
\caption{
Central potentials in the $^1S_0$ channel for three different quark masses
 in the quenched QCD simulations on a $(4.4 {\rm fm})^4$ lattice. }
\label{fig2}
\end{figure}

Fig.\ref{fig1} (upper panel) shows  the BS wave functions in  $^1S_0$ and $^3S_1$
channels for $\kappa=0.1665$.  The suppression of the wave function in
the region $r<  0.5$ fm indicate the existence of repulsion at short
distance.   Fig.\ref{fig1} (lower panel) shows  the reconstructed  NN potentials
for $\kappa=0.1665$,  i.e., the effective central potentials 
 for  $^1S_0$ and    $^3S_1$   channels.    See
Table~\ref{table},  for the  values of  the  non-relativistic energies
$E_q \equiv q^2/m_{\rm N}$ in Eq.(\ref{eq:naive_pot}).

We show
 the  NN  potentials  for  three  different quark  masses  in  $^1S_0$
 channel in Fig. \ref{fig2}.
 As the quark  mass decreases, the repulsive core at short
distance  is  enhanced  rapidly,  whereas  the  attraction  at  medium
distance is  modestly enhanced.   This indicates that  it is  important to
perform the lattice QCD calculations  for the lighter quark mass region in
order to make quantitative comparison of our results with the 
 observables in the real world.

Although there exist both attraction and repulsion,
 the net effect  of our potential is  attractive at low energies
  as shown  by the scattering lengths calculated from the 
  L\"uscher's formula  in Table \ref{table_a0} \cite{luscher}.
The attractive nature of our potential is 
qualitatively  understood by the Born  approximation formula for  the scattering
length
$
  a_0
  \simeq
  -m_N\int V_{\rm C}(r) r^2 dr.
$
Owing to the volume factor $r^2 dr$, the attraction at medium distance
 can overcome the repulsive core at short distance.
 However, there is  a considerable discrepancy  between
 the scattering   lengths in Table \ref{table_a0}   and   the  empirical   values,   
$a_0^{\rm(exp)}(^1S_0)$ $\sim$  $20$ fm and $a_0^{\rm(exp)}(^3S_1)\sim -5$ fm.
   This is attributed 
to the heavy quark masses employed in our simulations.

 If we  can get closer  to the physical  quark mass, there  appears an
 ``unitary region" where the  NN scattering length becomes singular 
  and    changes    sign as a    function    of    the    quark    mass   
 \cite{kuramashi96,meissner03,savage06}.  The   singularity is  associated with 
 the  formation of the di-nucleon bound state.
  Because of this, the NN scattering length becomes a highly non-linear function
   of the light quark mass.   One should note that the NN potential
   changes smoothly even in the unitary region 
   in contrast to the scattering length.  This is one of the 
    reasons why the NN potential
    is much more  appropriate  quantity to be examined on the lattice 
    instead of the NN scattering length.  

\begin{table}[t]
\begin{center}
\begin{tabular}{llll}
\hline
$\kappa$ &  $m_{\pi}$ [MeV] & $a_0(^1S_0)$ [fm] & $a_0(^3S_1)$ [fm] \\
\hline
0.1640   & 732.1(4) & $\ \ 0.115(26)$ & $\ \ 0.140(31)$\\
0.1665   & 529.0(4) & $\ \ 0.126(25)$  & $\ \ 0.140(31)$\\
0.1678   & 379.7(9) & $\ \ 0.159(66)$  & $\ \ 0.252(104)$\\
\hline
\end{tabular}
\end{center}
\caption{Scattering lengths obtained from the L\"uscher's formula \cite{luscher}
 in the spin-singlet and spin-triplet channels for different quark masses. }
\label{table_a0}
\end{table}
    
\section{Summary and concluding remarks}
\label{sec:summary}

 In this article, we have outlined
  the basic notion of the nucleon-nucleon potential and its field-theoretical 
  derivation from the 
  equal-time Bethe-Salpeter amplitude. 
  Such a formulation allows us to extract the potential
  between extended objects by using the lattice QCD simulations.
  The central part of the NN potential  at low energies was
   obtained in lattice QCD simulations with quenched approximation.
     It was found that
  the NN potential calculated on the lattice at low energy shows
   all the basic features 
  expected from the empirical NN potentials determined from
   the NN scattering data; attraction
   at  long and medium distances and the repulsion at short
  distance.  This is the first step toward the understanding 
  of atomic nuclei  from the fundamental law of the strong
   interaction, the quantum chromodynamics.

  There are a number of  directions to be explored  on the basis of our approach: 
\begin{itemize}
  \item[1.] Energy dependence of the NN potential 
   in Eq.(\ref{eq:naive_pot}) should be studied
   to test the non-locality of the potential $U(x,x')$ and
    the  validity of its derivative expansion.
  This is currently under investigation 
   by changing the spatial  boundary condition of the fermion field
   \cite{aoki_lat08}.
  \item[2.] The tensor force, which mixes the states with different orbital
   angular momentum by two units, is a unique feature of the 
    nuclear force and plays an essential role for 
    the deuteron binding.  This is also under investigation by
     projecting out
   the $^3S_1$ component and $^3D_1$ component separately from the
   exact two-nucleon wave function with $J=1$ on the lattice 
   \cite{Ishii_chiral07,Ishii_lat08}.
  The spin-orbit force, which is known to be strong at short
  distances in empirical NN force, should be also studied. 
 \item[3.] Three nucleon force is thought to play important roles in
 nuclear structure and also in the equation of state 
 of  high density matter.  Since the experimental information is
 scarce, simulations of the three nucleons  on the lattice 
 combined with appropriate generalizations of the formulas in Section 2 may
 lead to the first principle determination of the three
 nucleon potential in the future.  With these generalizations of the
  present approach, one may eventually make a firm link between QCD
   and the physics of nuclear structure  \cite{nuclear_app}.
  \item[4.] The hyperon-nucleon (YN) and hyperon-hyperon (YY)
   potentials are essential for understanding the properties
    of hyper nuclei and the hyperonic matter inside the
     neutron stars.  However, the experimental data are
      very limited due to the short life-time of hyperons.  
 On the lattice,  NN, YN and YY interactions 
  can be treated in the same footing since the difference is
   only the mass of the strange quark. Recently, 
  the $\Xi$N potential \cite{Nemura_lat07,Nemura08}
   and the $\Lambda N$ potential \cite{Nemura_lat08} are examined
   as a first step toward systematic derivation of the hyperon
    interactions.
  \item[5.] To compare the NN potential on the lattice with
    experimental observables, it is necessary to 
    carry out full QCD simulations which take into account the  dynamical quark loops.
   Study of the nuclear force with the use of the  
    full-QCD configurations generated by PACS-CS
    Collaboration \cite{Kuramashi:2007gs} is currently under investigation \cite{aoki_lat08,Ishii_lat08,Nemura_lat08}.
\end{itemize}

\section*{Acknowledgments}

This research was partly supported by the Ministry of Education, Culture, Sports, Science and Technology, 
 Grant-in-Aid Nos. 18540253, 19540261 and 20340047. Numerical simulations were supported 
 by the Large Scale Simulation Program No.07-07 (FY2007)
 of High Energy Accelerator Research Organization (KEK).
We are grateful for authors and maintainers of {\tt CPS++}\cite{cps},
of which a modified version is used for measurement done in this work.

\end{document}